# Community Lessons from builders of infrared multi-object spectrographs

Solange V. Ramirez*[a], Maren Cosens[a], Nicholas P. Konidaris II[a], Francisco Garzón[b], Brian McLeod[c], Suzanne Ramsay[d], Leon Aslan[e], Robert Barkhouser[f], Christoph Birk[a], Julia Brady[g], Tyson Hare[a], Stephen C. Hope[h], Charlie Hull[a], Karim Kaismoune[a], Daniel D. Kelson[a], Gerrad Killion[a], Alicia Lanz[i], Jacob McCloskey[j], Andrew B. Newman[a], Gwen C. Rudie[a], William Schoenell[a], Patricio Schurter[k], Stephen A. Smee[h], Jason E. Williams[a]

[a] Carnegie Institution for Science, The Observatories, 813 Santa Barbara St., Pasadena, CA 91107, USA; [b] Instituto de Astrofísica de Canarias; [c] Smithsonian Astrophysical Observatory; [d] European Southern Observatory; [e] Turion Space; [f] LCS Optics; [g] The Ohio State University; [h] Johns Hopkins University; [i] Capella Space; [j] Applied Research Laboratory at Pennsylvania State University; [k] Giant Magellan Telescope

## ABSTRACT

We present a summary of lessons learned from our community on the design and construction of infrared multi-object spectrographs. As members of the Magellan Infrared Multi-Object Spectrograph (MIRMOS) development team, we were motivated to learn about the most common troubles faced by our colleagues and incorporate measures in our design and assembly plan to address known issues. The lessons are compiled as a result of interviews we carried out with members of the instrument development teams for several multi-object infrared spectrographs. We found considerable overlap in the lessons learned for building this type of instrument; many of which relate to the cryogenic operation of the instruments. In this paper, we will present the compilation of general and specific lessons which we found relevant for the successful development of MIRMOS or any other future cryogenic instruments.

## 1. INTRODUCTION

We present a compilation of feedback and lessons learned from teams that developed previous multi-object infrared spectrographs, with the goal of informing the design and construction of the Magellan Infrared Multi-Object Spectrograph (MIRMOS). MIRMOS, destined for the 6.5-meter Magellan Telescopes in Chile, will support a wide range of science including extragalactic, cosmological, and exoplanetary research. It will deliver simultaneous R~3700 spectra across the Y, J, H, and K infrared bands using a complex optical path with a dichroic tree, VPH gratings, multiple cameras and detectors, a configurable slit unit producing up to 92 slits over a 13’x3’ field of view (FoV), and an integral field unit offering a ~18″×26″ FoV. The design is described in detail in [1], [2], [3], [4]. To aid in MIRMOS’s development, two members of the MIRMOS team reviewed the technical details of six similar instruments and conducted interviews with key instrument builders on various topics of interest. Despite variations among instruments, common lessons emerged, particularly concerning cryogenic operation and testing. The lessons learned during these activities have been critical for working on the design and assembly plan of MIRMOS and are likely useful for any team working on astronomical instruments, particularly those operating at cryogenic temperatures.

## 2. LESSONS GATHERING ACTIVITY

The main objective of gathering lessons for the MIRMOS instrument team was to identify specific areas or topics we might be missing to address during our design process and assembly, integration and testing plans. It was decided early on that the process of gathering information should be done in a systematic way by selecting instruments with similar functionality to MIRMOS and compiling a set of topics to address with the team members of each of the selected instruments.

*sramirez@carnegiescience.edu

The instruments selected to include in our study follow a simple criterion: they must be spectrographs with multi-object capability, they must generate infrared spectra covering a wavelength range of ~1um to 2.5um at moderate spectral resolution (R ~ 1000-3000), and they must operate on a telescope with similar aperture as the 6.5m Magellan telescope. The selected instruments were:

- MMIRS – MMT and Magellan InfraRed Spectrograph[5]
- FLAMINGOS-2 [6], [7]
- LUCI – LBT Utility Camera in the Infrared [8], [9]
- KMOS – K-band Multi Object Spectrograph [10]
- MOSFIRE – Multi-Object Spectrometer For InfraRed Exploration [11]
- EMIR - Espectrógrafo Multiobjeto Infra-Rojo [12]

Important features and technologies used in the instruments selected for this study are listed in Table 2-1, together with MIRMOS' design characteristics for comparison.

*Table 2-1: Features and technologies for selected instruments*

| Parameters | MMIRS | FLAMINGOS-2 | LUCI | KMOS | MOSFIRE | EMIR | MIRMOS |
|---|---|---|---|---|---|---|---|
| Telescope | Magellan | Gemini | LBT | VLT | Keck | GTC | Magellan |
| Aperture (m) | 6.5 | 8.0 | 8.4 | 8.4 | 10 | 10.4 | 6.5 |
| Wavelength range (total) | 0.9-2.4um | 0.9-2.4um | 0.9-2.5um | 0.8-2.5um | 0.9-2.4um | 0.9-2.5um | 0.9-2.4um |
| Resolution | 800-3000 | 300-1300 | 300-30,000 | 3300-3800 | 3300 | 3500-5000 | 3700 |
| MOS FOV | 6.9'x6.9' | 6.2' | 4'x4' | 7.2' | 6.1'x6.1' | 6.67'x4' | 13'x3' |
| Detectors | H2RG (1) | H2RG (1) | H2RG (1) | H2RG (1) | H2RG (1) | H2RG (1) | H2RG (4) |
| Detector Temp | 77K | 80K | 77K | 80K | 77K | 40K | 80K |
| Optics Temp | 77K | 110K | | 140K | <130K | 80K | 120K |
| Cooling system | LN2 | Cryocooler (CTI) | Cryocooler (Sumitomo) | Cryocooler (Leybold) | Cryocooler (CTI) | Cryocooler | Cryocooler |
| Slit mechanism | masks | masks | masks | arms | CSU | CSU | CSU |
| # slits | <= 60 | Up to 150 (length dependent) | 100 | 24 IFU arms | 46 | 53 | 92 |
| IFU slices | - | - | - | 0.2"x2.8" (14) | - | - | 0.84"x26" (23) |
| Dispersers | Grism/VPH | Grism | VPH | grating | grating | prism+grating | VPH |
| Commissioned | 2009 | 2013 | 2009 | 2013 | 2012 | 2016 | 2029 |

The process of information gathering included researching the selected instrument designs using information in the literature and instrument websites and contacting one instrument builder team member to carry out an interview. The interview preparation included identification of critical areas of inquiry and the creation of an outline of topics to address the interviewees. The main areas were understanding the primary issues or successes encountered during assembly, integration, and testing, as well as discussion of key cryogenic technologies and detectors.

The interviews with the instrument builders occurred over a period of a few months, which was set by the availability of our colleagues and interviewees. Each interview lasted about one hour, where notes were taken simultaneously by two MIRMOS team members conducting the interviews. The notes were checked for completeness and organized by topics. After all the interviews were conducted, the information was compiled into general categories and specific items. The general category items were issues encountered and/or raised by more than one of the interviewees and denote overlap in the lessons learned by different instrument building teams.

## 3. LESSONS LEARNED

During the interview process, we realized that there was considerable overlap in the lessons learned for building multi-object infrared spectrographs. Many of those lessons relate to the fact that mechanisms and optics need to operate at cryogenic temperatures. There is consensus among the instrument builders on these general lessons learned.

**Design Considerations**

Most of the instruments investigated made use of some type of cryocooler, as is the strategy for cooling MIRMOS. As this was an area being designed at the time we inquired with the interviewees about their strategy to deal with the vibration of these coolers and its success. In all cases vibration damping was incorporated in the design of the mounting system for these coolers. In general, passive damping designs such as those utilizing bellows were sufficient to meet the image quality requirements of the instruments. However, in cases with strict stability requirements for either the instrument or facility (e.g. for interferometry) this was not always sufficient and more detailed design consideration may be necessary.

For instruments which see multiple gravity vectors while observing, flexure can be a challenge. It was important for the design teams to perform finite element analysis (FEA) of the expected flexure, but this has often had large margins of error typically in the direction of underestimating the flexure. This may partly be due to the need to simplify the instrument models to speed up computation, for example by including point masses rather than distributed loads or modeling individual components as a larger monolithic structure. Given the need for at least some simplifying assumptions, it was often repeated that the design of a flexure compensation system must include large margins to account for these model deficiencies. Once built, testing the amount of flexure in the lab where mitigation of excessive flexure can still be incorporated is key.

**Assembly & Testing**

As we were conducting these interviews, we were also writing the alignment plan for the many optical components of MIRMOS. As the initial alignment must be done warm while it will be used cold, this was an area that was repeatedly emphasized as requiring careful procedures and testing plans. Particularly, lenses with aspheres can be challenging to align. It was recommended to break this alignment and testing into smaller subsystems to measure performance and focus before incorporation in the full instrument to improve the ability to diagnose problems.

Aside from the testing of optics, all these instruments (as well as MIRMOS) had cryogenic mechanisms of varying complexity which required (or would have benefitted from) extensive testing. It was emphasized that this should be done at the component and system level, both with preliminary warm checks and robust testing at the instrument operating temperature. It was recommended to test exactly what will be used as seemingly small changes can have a significant impact on the reliability and lifetime of a mechanism, particularly when dealing with cryogenic operation.

A common refrain was that allowing time for sufficient testing is key to a smooth commissioning process as well as the success of the instrument. There can often be a tendency to descope testing plans to meet observatory deadlines and ship on time. However, this can lead to problems going undiscovered until commissioning and performing additional testing and modifications can be a challenge at the observatory. It was often recommended to allow sufficient margin and schedule contingency for testing activities and to delay shipment rather than skipping planned tests.

**Software Development**

There was wide consensus that software development tasks are generally underestimated and lag behind other engineering tasks. The general lesson here is to start early on software development to provide basic tools for testing and commissioning. One example is observing planning tools, which can be crucial for success during commissioning as they enable smooth and realistic on-sky testing. Another example is having working data reduction pipelines, which can facilitate the data analysis during early on-sky activities. The general sentiment was that teams which arrived at the telescope with a working instrument accompanied by observing planning tools and data reduction pipelines had more successful and faster commissioning phases than the ones lacking such software tools.

Additionally, it was recommended to have lab testing results well organized and accessible during on-site commissioning as this provides a means to understand instrument behavior when integrated with the respective telescopes. For example, on-sky positioning can be time consuming even when coordinate transformations from celestial to focal plane coordinates are optically modeled and/or estimated at the lab.

**Management**

There is a tendency to underestimate project management and software development efforts, and it highlighted the importance of working on a well-balanced team where all disciplines are represented and working together. This particular issue may come from the desire to develop instruments using a cost-effective model. It was recommended to determine the balance of engineering disciplines within the instrument team depending on the size and complexity of the instrument, with emphasis on paying attention to the traditionally underestimated engineering fields.

One area that was particularly discussed was communication and planning with the observatories. It was widely recommended to engage the observatories early and often. It is important to establish the instrument's functionality needs in terms of mass, power, software, and personnel. Mechanical and electronic interfaces are particularly important to be agreed upon and documented, as well as software requirements and upgrades from the observatory side, installation needs in terms of time and personnel, and storage availability and support.

**Specific Issues**

Here are some specific lessons that came up during the interviews and are relevant and interesting to consider:

1. Test warm motors impact on background emission
2. Unplanned warm-up testing of failure modes (without optics). Some failures should not rely on procedures, incorporate fail safes:
    a. Control on power supplies to shut down on a temperature threshold (thermostat in series that cut power)
3. Include pressure relief valve in your cryostats
4. Encoders preferred over detents
5. Include more test points than you think will be needed: temperature sensor, measuring reference points
6. For fast camera detector alignment, preferable to have a motorized focus and tip-tilt mechanism for faster alignment
7. Keep mechanisms as simple as possible, with fewer subcomponents which can fail.
8. For under-sampled point sources: there are patterns in the continuum, which are difficult to reduce in the data.
9. Communication of requirements across subgroups and ensuring compliance of specifications with testing procedures
10. Instrument mass: include margin early on
11. Opticentric tip on aspheres: Use local radius of curvature at the center and small beam diameter
12. Using legacy designs should involve analysis with original teams to understand the issues
13. Develop Concept of Operations (ConOps) early and incorporate operation feature at the component level software
14. Organize data collected at the lab and scripts for their analysis and have it available during commissioning.

## 4. LESSONS APPLIED TO THE MIRMOS DESIGN AND PLANNING

The compiled lessons learned had a positive impact on the MIRMOS team. After the final compilation of lessons, we had several meeting sessions where the findings were discussed. We determined that there were areas where our team was already embracing some of our colleagues' recommendations, for example in the mechanism testing, instrument flexure modeling and compensation, and definition of ConOps and identification of observing planning tools to be developed in parallel with other design activities. Other findings provided more context and rationale to advocate for a staffing profile that will allow our team to satisfy our software development and project management needs. Lastly, it gave our team specific information to include during the design and construction of MIRMOS to ensure that our limited resources were used in addressing high return items. For example, flexure modeling was identified as a critical activity with potentially large margins of uncertainty. To address this, we further explored the assumptions and simplifications made in our analysis to provide the most robust model realistically possible of the flexure and its impact on image motion along with robust evaluation of the margin required on compensation. Smaller adjustments were also made such as adding more mounting points for temperature sensors to allow for their use in troubleshooting efforts during the testing phase (but removed during operation).

## 5. CONCLUSIONS

The activity of collecting lessons learned from the community has yielded important information to share within our MIRMOS team and many other instrument builders. Key recommendations include thorough warm- and cold-temperature testing of all mechanisms; early development of control software and data-reduction pipelines; clear alignment procedures with incorporated cold testing; and careful modeling and measurement of instrument flexure. Additional guidance emphasizes vibration-aware cooling design, early coordination with the observatory, and balanced staffing. Specific advice includes testing warm motor impacts, adding fail-safes and pressure-relief valves, increasing test-point availability, simplifying mechanisms, accounting for under-sampling effects, tracking requirements across teams, budgeting mass margins, and consulting original designers when reusing legacy designs.

## ACKNOWLEDGEMENTS

We are extremely grateful to all our colleagues who agreed to be interviewed for this study and did so honestly and openly: Stephen Eikenberry, Francisco Garzón, Nick Konidoris, Brian McLeod, Suzanne Ramsay, and Walter Seifert. We thank the entire MIRMOS team and our collaborators at CSEM, Teledyne, and INAF. This material is based on substantial funding from Carnegie Science. This research was funded by the Heising-Simons Foundation through grant 2021-2614. This material is based upon work supported by the National Science Foundation under Grant No. 2206374. We recognize generous support from the Ahmanson Foundation. The Mt. Cuba Astronomical Foundation funded MIRMOS.

## REFERENCES

[1] Konidaris, N. P., Rudie, G. C., Newman, A. B., et al "The Magellan Infrared Multiobject Spectrograph project", Proc. SPIE 11447, Ground-based and Airborne Instrumentation for Astronomy VIII, 114471E (13 Dec 2020); https://doi.org/10.1117/12.2561171

[2] Konidaris, N. P., Rudie, G. C., Newman, A. B., et al "The Magellan infrared multi-object spectrograph project: 2022 update", Proc. SPIE 12184, Ground-based and Airborne Instrumentation for Astronomy IX, 1218415 (29 Aug 2022); https://doi.org/10.1117/12.2626995

[3] Konidaris, N. P., Rudie, G. C., Newman, A. B., et al "The Magellan infrared multi-object spectrograph project: 2024 update", Proc. SPIE 13096, Ground-based and Airborne Instrumentation for Astronomy X, 130960T (18 Jul 2024); https://doi.org/10.1117/12.3019306

[4] Cosens, M., et al; this proceedings

[5] McLeod, B., Fabricant, D., Nystrom, G., et al. “MMT and Magellan Infrared Spectrograph”, PASP, 124, 1318

[6] Eikenberry, S., Elston, R., Raines, S. N., et al. “FLAMINGOS-2: the facility near-infrared wide-field imager, and multi-object spectrograph for Gemini”, Proc. SPIE 5492, Ground-based Instrumentation for Astronomy, (30 Sept 2004); https://doi.org/10.1117/12.549796

[7] Eikenberry, S., Bandyopadhyay, R., Bennet, J. G., et al. “FLAMINGOS-2: the facility near-infrared wide-field imager and multi-object spectrograph for Gemini", Proc. SPIE 8446, Ground-based and Airborne Instrumentation for Astronomy IV, 84460I (24 Sept 2012); https://doi.org/10.1117/12.925679

[8] Seifert, W., Appenzeller, I., Baumeister, H., et al. "LUCIFER: a multimode NIR instrument for the LBT", Proc. SPIE 4841, Instrument Design and Performance for Optical/Infrared Ground-based Telescopes, (7 Mar 2003); https://doi.org/10.1117/12.459494

[9] Mandel, H. G., Appenzeller, I., Seifert, W., et al. "LUCIFER status report: Summer 2006", Proc. SPIE 6269, Ground-based and Airborne Instrumentation for Astronomy, 62693F (11 Jul 2006); https://doi.org/10.1117/12.671067

[10] Sharples, R., Bender, R., Berbel, A., A., et al. "Recent progress on the KMOS multi-object integral-field spectrograph for ESO VLT", Proc. SPIE 7735, Ground-based and Airborne Instrumentation for Astronomy III, 773515 (15 Jul 2010); https://doi.org/10.1117/12.856125
[11] McLean, I. S., Steidel, C. C., Epps, H., et al. "Design and development of MOSFIRE: the multi-object spectrometer for infrared exploration at the Keck Observatory", Proc. SPIE 7735, Ground-based and Airborne Instrumentation for Astronomy III, 77351E (20 Jul 2010); https://doi.org/10.1117/12.856715
[12] Garzón, F., Abreu D., Barrera S., et al. "EMIR: the GTC NIR multi-object imager-spectrograph", Proc. SPIE 6269, Ground-based and Airborne Instrumentation for Astronomy, 626918 (28 Jun 2006); https://doi.org/10.1117/12.671302